\documentclass{ws-procs9x6}
\newcommand{\up}{\uppercase}
\def\rref#1{(\ref{#1})}
\def\be{\begin{equation}}
\def\ee{\end{equation}}
\def\bea{\begin{eqnarray}}
\def\eea{\end{eqnarray}}
\def\ba{\begin{array}}
\def\ea{\end{array}}

\def\sg{\sigma}
\def\um{\frac{1}{2}}
\def\b{\beta}
\def\da{\delta}
\def\a{\alpha}
\def\om{\omega}
\def\Om{\Omega}
\def\ga{\gamma}

\def\La{\Lambda}
\def\la{\lambda}
\def\ep{\epsilon}
\def\IR{{\hbox{{\rm I}\kern-.2em\hbox{\rm R}}}}
\def\IZ{{\hbox{{\rm I}\kern-.2em\hbox{\rm Z}}}}
\def\IC{{\hbox{{\rm I}\kern-.2em\hbox{\rm C}}}}
\def\II{{\hbox{{\rm I}\kern-.2em\hbox{\rm I}}}}
\def\IT{{\hbox{{\rm I}\kern-.2em\hbox{\rm T}}}}

\def\delbar{\,\overline{\mathop{\!\nabla\!}}\,}

\begin{document}

\title{Some Applications of the ADM Formalism\footnote{\uppercase{W}ork
supported by the \up{I}stituto \up{N}azionale di \up{F}isica \up{N}ucleare
(\up{INFN}) of \up{I}taly and the \up{I}talian \up{M}inistero
dell'\up{U}niversit\`a e della \up{R}icerca \up{S}cientifica e \up{T}ecnologica
(\up{MIUR}).}}

\author {J.\ E.~N{\sc elson}}

\address{Dipartimento di Fisica Teorica, Universit\`a degli Studi
       di Torino\\
 and Istituto Nazionale di Fisica Nucleare, Sezione di Torino\\
       via Pietro Giuria 1, 10125 Torino, Italy\\
       E-mail: nelson@to.infn.it}

\maketitle

\abstracts{The ADM Formalism is discussed in the context of $2+1$--dimensional
gravity, uniting two areas of relativity theory in which Stanley Deser has been
particularly active. For spacetimes with topology $\IR \times T^2$ the
partially reduced and fully reduced ADM formalism are related and quantized,
and the role of "large diffeomorphisms" (the modular group) in the quantum
theory is illustrated.}

\section{Introduction}\label{sec1}
Over forty years ago Arnowitt, Deser and Misner (ADM) studied the 
$3+1$--decomposition of general relativity, its initial value problem, the 
dynamical structure of the field equations and calculated the Hamiltonian
\cite{adm}. This extraordinary piece of work has become a fundamental
ingredient of modern relativity theory. It is now regularly taught as an
integral part of relativity courses, and usually occupies at least a chapter
in relativity textbooks. The problem had actually been considered previously
by Dirac \cite{dir1} who applied his theory of constrained systems \cite{dir2}
to the gravitational field. But Dirac's treatment was incomplete and in a
particular gauge. 

In this Section I briefly summarise the ADM results,and in Section \ref{sec2}
discuss the main differences between the $3$ and $4$ dimensional theories. In 
Section \ref{metadm} the second--order, partially reduced, ADM formalism, for
spacetimes of topology $\IR \times T^2$ is reviewed, and I show how in
principle the system can be quantized. In Section \ref{triadm} the
first--order fully reduced holonomy approach is presented. In Section
\ref{sec5} the two approaches are related, both classically and quantum
mechanically, using  the action of the modular group, or ``large
diffeomorphisms - those that remain after ADM reduction. 

The $3+1$--decomposition of the Einstein--Hilbert action calculated by ADM is  
\be 
I_{\hbox{\scriptsize \it Ein}} = \int\!d^4x
\sqrt{-{}^{\scriptscriptstyle(4)}\!g}\> {}^{\scriptscriptstyle(4)}\!R  
= \int dt\int\nolimits_\Sigma d^3x \bigl(\pi^{ij}{\dot g}_{ij} 
- N^i{\mathcal H}_i - N{\mathcal H}\bigr), 
\label{ein} 
\ee 
where spacetime is of the form
$\IR\times\Sigma$, and time runs along $\IR$. In \rref{ein} the metric has been
decomposed as 
\be 
ds^2 = N^2dt^2 - g_{ij}(dx^i + N^i dt)(dx^j + N^j dt) \qquad i,j = 1,2,3 
\label{met} 
\ee 
and $\pi^{ij} = \sqrt{g}\,(K^{ij} - g^{ij}K)$, where $K^{ij}$
is the extrinsic curvature of the surface $\Sigma$ labeled by $t={\rm
const.}$\footnote{This is standard ADM notation: $g_{ij}$ and $R$ refer to the
induced metric and scalar curvature of a time slice, while the spacetime metric
and curvature are denoted ${}^{\scriptscriptstyle(4)} \!g_{\mu\nu}$ and
${}^{\scriptscriptstyle(4)}\!R$.} In \rref{met} the lapse $N^i$ and shift $N$
functions are related to the non-dynamical components of $g_{ij}$ and their
variation in \rref{ein} leads to the supermomentum and super-Hamiltonian
constraints on $g_{ij}$ and $\pi^{kl}$.  
\be 
{\mathcal H}_i = -2\nabla_j{\pi^j}_{i}\,= 0, \qquad 
{\mathcal H} = \frac{1}{\sqrt{g}}\,g_{ij}g_{kl}(\pi^{ik}\pi^{jl}
-\um\pi^{ij}\pi^{kl}) -\sqrt{g} R = 0 
\label{constr} 
\ee 
where $\nabla_j$ is the covariant derivative for the connection compatible with
$g_{ij}$, and indices are now raised and lowered with $g_{ij}$. The ${\mathcal
H},{\mathcal H}_i$ in \rref{constr} are  non-polynomial in $g_{ij}$ and
$\pi^{kl}$ and involve $(g_{ij})^{-1}$. They are  directly proportional to the
components $G^{o\mu}$ of the Einstein tensor defined by
\be G^{\mu\nu}~=\frac{\da I_{\hbox{\scriptsize \it Ein}}}{\da
g_{\mu\nu}} = R^{\mu\nu} -\um R g^{\mu\nu}
\label{eint}
\ee  
so finding a solution to \rref{constr} would correspond to finding a general
solution of Einstein's equations (the other components of \rref{eint} are zero 
by the Bianchi identities ${G^{\mu\nu}}_{|\nu} = 0$). 

The constraints \rref{constr} generate, through the Poisson brackets obtained 
from \rref{ein}
\be
\{g_{ij}(x),\pi^{kl}(y)\}= \um (\da^{k}_{i}\da^{l}_{j}+\da^{k}_{j}\da^{l}
_{i})\da^{3}(x-y) \label{pb} 
\ee 
three-dimensional diffeomorphisms in $\Sigma$, and the time development of the
variables $g_{ij}(x),\pi^{kl}(y)$.

One can ask what effect the constraints \rref{constr} would have when {\it
applied} on wave functions $\psi(g)$. If the brackets \rref{pb} are
represented by letting the momenta  $\pi^{ij}$ act by differentiation 
\be
\pi^{ij}(x) \sim \frac{\da}{\da g_{ij}(x)} 
\label {pirep} 
\ee
and the metric components $g_{ij}(x)$ by multiplication, the supermomentum 
constraint ${\mathcal H}_i \psi(g) = 0$ is easy to interpret, since
\be
\int\nolimits_\Sigma d^3x \bigl(N^i{\mathcal H}_i\bigr)\Psi(g) = \da\Psi(g) = 
\frac{\da\Psi(g)}{\da g_{ij}(x)}(\nabla_i N_j + \nabla_j N_i) 
\label{hi}
\ee
and implies that one should identify wave functions of metrics $g_{ij}$ and
$\tilde {g} _{ij}$ when they differ as
\be
\tilde {g} _{ij} = g_{ij} +\nabla_i N_j+\nabla_j N_i.
\label{hipsi}
\ee
But \rref{hipsi} is a Lie derivative, or coordinate transformation, 
in the spatial surface $\Sigma$, so the supermomentum constraint reflects the
freedom to choose the $3$ spatial coordinates on $\Sigma$. The space of metrics 
\rref{hipsi} with $\tilde {g} _{ij}$ identified with $g_{ij}$ was named 
superspace in 1963 by Wheeler \cite{mtw}.

The super-Hamiltonian constraint ${\mathcal H}\psi(g) = 0$ (also known as the 
Wheeler DeWitt equation \cite{wdw}) is much harder to interpret, and alone 
does {\it not} generate the
dynamics, or time reparametrization invariance, of wave functions $\psi(g)$.
Instead one needs to use the full Hamiltonian, namely the combination
\be
\int\nolimits_\Sigma d^3x \bigl(N^i{\mathcal H}_i + N{\mathcal H}\bigr).
\ee
The gravitational field in $4$ spacetime dimensions has correctly (for a
massless field) $2$ independent degrees of freedom per spacetime point. This is
most easily seen by noting that the induced metric  of a time slice $g_{ij}$ 
has $6$ independent components, and there are the $4$ constraints \rref{constr}. 

\section{$2+1$--Dimensional ADM Decomposition}\label{sec2}
In $2+1$ spacetime dimensions the description of Section \rref{sec1} is
essentially identical, apart from a factor of $\um$ in the super-Hamiltonian
\rref{constr}. The counting of 
degrees of freedom is, however, quite different. There are in fact zero degrees
of freedom, and this can be seen in several ways. The simplest is perhaps to
note that now the induced matric $g_{ij}, i, j=1,2$ has only $3$ independent
components, but there are $3$ constraints ${\mathcal H}=0,  {\mathcal H}_i = 0$
analogous to \rref{constr}. Alternatively, since the Weyl tensor vanishes in
$3$ dimensions (but not in $4$, see \cite{jac}), it follows that the full
Riemann curvature tensor $R_{\alpha\beta\mu\nu}$ can be decomposed uniquely in
terms of only the Ricci tensor  $R_{\mu\nu}$ the scalar curvature $R$ and the
metric tensor $g_{\mu\nu}$ itself.
\bea
R_{\la\mu\nu k} =g_{\la\nu} R_{\mu k} - g_{\mu\nu} R_{\la k} 
- g_{\la k} R_{\mu\nu} + g_{\mu k} R_{\la\nu}\nonumber\\
+ \um R(g_{\mu\nu} g_{\la k} - g_{\la\nu} g_{\mu k})
\label{curv}
\eea
In fact in $d$ dimensions $R_{\alpha\beta\mu\nu}$ has $\frac{d^2(d^2-1)}{12}$ 
independent degrees of freedom and $R_{\mu\nu}$ has $\frac{d(d+1)}{2}$. These 
coincide when $d=3$. In terms of the Einstein tensor $G^{\a\b}$ (equation
\rref{eint}) the decomposition \rref{curv} is
\be 
R_{\la\mu\nu k}\ =\ \ep_{\la\mu\b}\ \ep_{\nu k\a}\ G^{\a\b} \label{curvein}
\ee 
so that when Einstein's vacuum equations $G^{\a\b}= 0$ are satisfied, the full 
curvature tensor (all components) are zero, i.e. $R_{\la\mu\nu k}\ = 0 $
and spacetime is flat. Thus vacuum solutions of Einstein's equations 
correspond to flat spacetimes, and there are no local degrees of freedom.

It is possible, however, to solve the field equations and introduce some 
dynamics, in several ways. The first - developed extensively by Deser 
et al \cite{jac2} and others, is to add sources, or matter, thus creating local 
degrees of freedom. When Einstein's equations read
\be 
G^{\a\b}= T^{\a\b}\label{gt}
\ee
where $T^{\a\b}$ is the stress-energy tensor of the sources, the 
curvature \rref{curvein} is no longer zero, but is proportional, from 
\rref{gt}, to $T^{\a\b}$.

The second creates propagating massive gravitational modes by adding a 
topological term to the action, always possible in an odd number of 
dimensions \cite{jac}. For gravity in 3 dimensions, this is the Chern-Simons 
form
\be 
\int (\om^{ab}\wedge d\om_{ab}+\frac{2}{3}\om^{ac}\wedge\om^
{d}_{c}\wedge\om_{da})\label{cs}
\ee
where the components of the spin connection $\om^{ab}_{\mu}$ are to 
be considered as 
functionals of the triads $e^{a}_{\mu}$ by solving the torsion equation.
$$R^a = de^{a}-\om^{ab}\wedge e_{b}\ = 0 $$
with $e^{^a}_{\mu}e^{^b}_{\nu}\eta_{ab}=g_{\mu\nu}$.
Variation of \rref{cs} with respect to the metric tensor $g_{\mu\nu}$ gives 
the Cotton tensor
$$C^{\mu\nu}=g^{-\um}\ \ep^{\mu\la\b}D_{\la}\left( R^{\nu}_{\b}-\frac{1}{4} 
\da^{\nu}_{\b}R\right) $$
which is symmetric, traceless, conserved, and vanishes if the theory is 
conformally invariant. Therefore, adding the Chern-Simons term \rref{cs} to 
the three-dimensional scalar curvature action \rref{ein} with a constant 
factor $\frac{1}{\mu}$ leads to the field equations
$$G^{\mu\nu}+\frac{1}{\mu}C^{\mu\nu}=0$$
which can be transformed into
\be \Big(\square +\mu^{2}\Big)\ R_{\mu\nu}=\ {\rm terms\ in}\ \Big(
R_{\mu\nu}\Big)^{2}\label{lin} 
\ee
In the linearized limit the R.H.S. of \rref{lin} vanishes and it is shown in 
\cite{jac} that the solutions of \rref{lin} correspond to massive, spin $\pm 2$, 
particles.

A way to introduce global rather than local degrees of freedom in flat 
spacetime is to consider non trivial topologies. Recall that curvature is
defined by commutators of covariant derivatives, or, by parallel transport
around  non-collapsible curves i.e. curves which are not homotopic to the
identity. The change effected by parallel transport around closed curves of
this type is often called holonomy - and is used to characterise flat
spacetimes. A simple example is when the spatial surfaces are tori, i.e.
$\Sigma = T^2$ - then the meridian and parallel are clearly non-collapsible.
This will be discussed explicitly in Sections \rref{metadm} and \rref{triadm}.

\section{Second--Order, Partially Reduced ADM Formalism\label{metadm}} Here 
I summarise work by Moncrief \cite{mon} and Hosoya and Nakao  \cite{hosnak},
adding a cosmological constant $\La$. It is known that any two-metric $g_{ij}$ on
$\Sigma_g$, where  $\Sigma_g$ is a Riemann surface of genus $g$, is conformal
(up to a diffeomorphism) to a finite-dimensional family of constant curvature
metrics ${\bar g}_{ij}(m_\alpha)$,  
\be g_{ij} = e^{2\lambda}{\bar
g}_{ij}(m_\alpha) , \label{conf} 
\ee 
labelled by a set of moduli $m_\alpha, \a = 1.....6g-6$ (see Abikoff
\cite{abikoff}), and 
\be R(\bar g) = \ba{cc} 1& g=0\\ 0& g=1\\ -1&
g>1 \ea \label{cur} 
\ee 
A similar decomposition of the momenta $\pi^{ij}$ gives
\be \pi^{ij} = e^{-2\lambda}\sqrt{\bar g} \left( p^{ij}+ \um \bar
g^{ij}\pi/\sqrt{\bar g} + \delbar^iY^j + \delbar^jY^i - \bar g^{ij}\delbar_kY^k
\right) \label{mom} 
\ee
where $\delbar_i$ is the covariant derivative for $\bar g_{ij}$, indices are
now raised and lowered with $\bar g_{ij}$, and $p^{ij}$ - the momentum
conjugate  to ${\bar g}_{ij}$ - is transverse traceless with respect to
$\delbar_i$, i.e., $\delbar_i\, p^{ij} = 0$.

This decomposition uses York time \cite{york}, the mean (extrinsic) curvature 
$K = \pi/\sqrt{g} = T$, shown to be a good global coordinate choice in \cite{mon}. 

The supermomentum constraints now imply that $Y^i=0$, while the super-Hamiltonian 
constraint,
\be
{\mathcal H} = -\um \sqrt{\bar g}e^{2\lambda}(T^2 - 4\Lambda)
  + \sqrt{\bar g}e^{-2\lambda} p^{ij}p_{ij} + 2\sqrt{\bar g}\left[
  \bar\Delta\lambda - \um\bar R \right] = 0 ,
\label{hconstr}
\ee
reduces to a  differential equation for the conformal factor $\la$ as a 
function of $\bar g_{ij}, p^{ij}$ and $T$. For $g>1$ a solution of 
\rref{hconstr} always exists and the three-dimensional Einstein--Hilbert action
is 
\be
I_{\hbox{\scriptsize \it Ein}}
= \int dT \left( p^\alpha \frac{dm_\alpha}{dT} - H(m,p,T)\right)
\label{ein3}
\ee
where $p^\alpha$ are the momenta conjugate to the moduli $m_{\a}$ defined by
\be  
p^{\a} = \int _{\Sigma} d ^2 x ~~
\pi^{ij} \frac{\partial}{\partial m_{\a}} {\bar g}_{ij}.
\label{pa}
\ee 
and $H(m,p,T)$ is an effective, or reduced, ADM  Hamiltonian 
\be
H(m,p,T) = 
\int_\Sigma \sqrt g\,d^2x  = \int_\Sigma e^{2\lambda(m,p,T)}\sqrt{\bar g}\,d^2x 
\label{hred}
\ee
which represents the surface area at time $T$, with $\lambda(m,p,T)$ determined 
by \rref{hconstr}. The reduced ADM Hamiltonian \rref{hred} generates the 
$T = K$ or time development of $m_\a, p^\b$ through the Poisson brackets 
\be
\left\{ m_\alpha, p^\beta \right\} = \delta^\beta_\alpha.
\label{pbm}
\ee 
For $g = 1$ the modulus is the complex number $m = m_1 + im_2$ 
(with $m_2>0$), with momenta $p = p^1 + i p^2$ satisfying the Poisson brackets 
\be
\left\{ m, \bar p\right\} = \left\{ \bar m, p\right\} = 2 ,\quad
\left\{ m, p\right\} = \left\{ \bar m, \bar p\right\} = 0
\label{pbtor}
\ee
and 
\be
d\sigma^2 = m_2^{-1}\left| dx + mdy \right|^2 ,
\label{2met}
\ee
is the spatial metric for a given $m$ where $x$ and $y$ each have period $1$. 
The surface curvature \rref{cur} is zero and \rref{hconstr} is 
explicitly solved. The reduced ADM Hamiltonian \rref{hred} becomes
\be
H(m,p,T) = \left(T^2-4\Lambda\right)^{-1/2}\left[m_2{}^2 p\bar p\right]^{1/2}.
\label{hamtor}
\ee
One can recognise in \rref{hamtor} the square of the momentum with respect to 
the Poincar\'e (constant negative curvature) metric on the torus moduli space
\be
m_2{}^{-2} dm d\bar m .
\label{pmet}
\ee
Hamilton's equations for the motion of $m,p$ on the hyperbolic upper half plane
(Teichm\"uller space) using the reduced Hamiltonian \rref{hamtor} can be solved
exactly \cite{ez,fuj} and correspond to motion on a semicircle, a geodesic 
with respect to the metric \rref{pmet}.

This reduced phase space can, in principle, be
quantized by replacing the Poisson brackets \rref{pbm} with commutators,
\be
\left[ \hat m_\alpha, \hat p^\beta \right] =i\hbar\delta^\beta_\alpha
\label{comm}
\ee
representing the momenta as derivatives,
\be
{\hat p}^\alpha = \frac{\hbar}{i}\frac{\partial}{\partial m_\alpha} ,
\label{pop}
\ee
and imposing the Schr\"odinger equation
\be
i\hbar\frac{\partial\psi(m,T)}{\partial T} = \hat H\psi(m,T) ,
\label{schr}
\ee
where the Hamiltonian $\hat H$ is obtained from \rref{hamtor} by some
suitable operator ordering. With the ordering of \rref{hamtor}, the Hamiltonian
is 
\be
\hat H = \frac{\hbar}{\sqrt{T^2-4\Lambda}}\,\Delta_0^{1/2} ,
\label{hamop}
\ee
where $\Delta_0$ is the scalar Laplacian for the constant negative curvature 
moduli space with metric \rref{pmet}. Other orderings exist, but all consist of 
replacing $\Delta_0$ in \rref{hamop} by $\Delta_n$, the weight $n$ Maass 
Laplacian (see e.g. Carlip \cite{ord}).

This approach also depends on the arbitrary, albeit good, choice of $K =
\pi/\sqrt{g} = T$ as a time variable. It is not at all clear that a different
choice would lead to the same quantum theory.  

\section{First--Order Fully Reduced, ADM Formalism\label{triadm}}
The first-order, connection approach to (2+1)-dimensional gravity, in which the
triad one-form $e^a = e^a{}_\mu dx^\mu$ and the spin connection
$\omega^{ab}=\omega^{ab}{}_\mu dx^\mu$ are treated as independent variables was
inspired by Witten \cite{wit} (see also \cite{achu}) and developed by Nelson,
Regge and Zertuche \cite{NR0,NRZ,NR1,NR2}.  The three dimensional
Einstein-Hilbert action is
\be 
I_{\hbox{\scriptsize \it Ein}} = 
\int\- (d\omega^{ab}-{\omega^a}_d\wedge\omega^{db} +\frac{\Lambda}{3} 
e^a\wedge e^b)\wedge e^c\,\epsilon_{abc} ,
\qquad a,b,c=0,1,2. 
\label{eintri} 
\ee 
For $\Lambda < 0$ this action can be written (up to a total derivative) 
as\footnote{For $\Lambda \geq 0$ see e.g. the discussion in \cite{cn}}
\be
I_{\hbox{\scriptsize CS}} = - \frac{\a}{4}
\int(d~\Omega^{AB}-\frac{2}{3}\Omega^A{}_E\wedge\Omega^{EB}) \wedge\Omega^{CD}
\epsilon_{ABCD} , 
\label{cs3} 
\ee 
where $A,B,C.. = 0,1,2,3, \epsilon_{abc3}=-\epsilon_{abc}$, the tangent 
space metric is
$\eta_{AB}=(-1,1,1,-1)$  and the (anti-)de Sitter  $SO(2,2)$ spin connection
$\Omega^{AB}$ is  
\be {\Omega^A}_B=\left(
\begin{array}{cc} \omega^a{}_b& -\frac{e^a}{\a}\\ -\frac{e_b}{\a} & 0
\end{array} \right) . \label{bigom} 
\ee 
with $\Lambda = -\alpha^{-2}$. The canonical $2+1$-decomposition of \rref{cs3}
is  
\be I_{\hbox{\scriptsize \it Ein}} =
\int\!d^3x ({\Omega_i}^{AB}{{\dot{\Omega}}_j}^{CD}\ep_{ABCD}-
{\Omega_0}^{AB}R_{ABij}) \epsilon^{ij}. \label{eintri2}
\ee
In \rref{eintri2} the curvature two-form 
$R^{AB}=d~\Omega^{AB}-\Omega^{AC}\wedge\Omega_C{}^B$ has components $R^{ab}+\La
e^a \wedge e^b$ (proportional to the constraints \rref{constr}) and 
$R^{a3}=\frac{1}{\a}R^a$ (proportional to a rotation constraint $J_{ab}$ on the 
triads), where  
\be R^{ab} = d\omega^{ab} -
\omega^{ac}\wedge\omega_{c}{}^{b} ,\quad R^{a} = de^a -\omega^{ab}\wedge e_b 
\label{r2}  
\ee  
are the (2+1)-dimensional curvature and torsion. The
field equations (constraints) derived from the action \rref{eintri2} are simply
$R^{AB}=0$, and imply that the $SO(2,2)$ connection $\Om^{AB}$ is flat, or,
equivalently, from \rref{r2} that the torsion vanishes everywhere and that the 
curvature $R^{ab}$ is constant. They generate, through the Poisson brackets  
\be
\{{\Omega_i}^{AB}(x),{\Omega_j}^{CD}(y)\} =\frac{1}{2\a}\epsilon_{ij}
\epsilon^{ABCD}\da^{2}(x-y) . \label{pbom} 
\ee 
infinitesimal gauge and coordinate transformations $\da \Om^{AB} = Du^{AB}$ on 
the connections $\Omega^{AB}$.

Since the connection $\Omega^{AB}$ is flat, it can be written locally in terms
of an  $\hbox{SO}(2,2)$-valued zero-form $\psi^{AB}$ as $dG^{AB}=\Om^{AC}\,
G_C{}^B$. This sets to zero all the constraints $R^{AB}=0$ and is therefore a 
{\it fully} reduced ADM formalism in which the Hamiltonian is identically zero.
However, some global degrees of freedom remain, as can be seen by now taking
into account the non--trivial topology of the Riemann surface. For each path 
$\sg$ on $\Sigma$ define the holonomy (Wilson loop) 
\be  G_{\sg}^{AB} = \exp P \int_{\sg} \Om^{AB} 
\ee 
where $P$ denotes
path--ordered, and $G_{\sg}$ depends on the base (starting) point and the
homotopy class $\{\sg\}$ of $\sg$, and satisfies 
$G_{\sg \rho}=G_{\sg}G_{\rho}$.~Integrating the brackets \rref{pbom} along
paths $\rho,\sg$ with non--zero intersection gives
\be
\{G_{\rho}, G_{\sg}\} \neq 0 \label{pbg} 
\ee 
and this is the starting point for holonomy quantization. 

It is actually more convenient to use the spinor groups
$\hbox{SL}(2,\IR)\otimes \hbox{SL}(2,\IR)$ for $\hbox{SO}(2,2)$ (see \cite{NRZ}
for details). For each path $\sg$ we have
\be
G_{\sg}^{AB}\ga_B= S^{-1}(\sg)\ga^A S(\sg)
\ee
where $\ga^A$ are the Dirac matrices and 
$S=S^+ \otimes S^-, S^{\pm}~\ep~\hbox{SL}(2,\IR)$. Explicitly, if the paths 
$\rho, \sg$ have a single intersection then \cite{NRZ}
\bea
\{S^{\pm}(\rho)_{\a}^{\b}, S^{\pm}(\sg)_{\ga}^{\tau}\} &=& \pm s
(-S^{\pm}(\rho)_{\a}^{\b} S^{\pm}(\sg)_{\ga}^{\tau}+2S^{\pm}(\rho^{\ }_{3} \sg^{\ }_{1})_{\a}^{\tau}
S^{\pm}(\sg^{\ }_{3}\rho^{\ }_{1})_{\ga}^{\b}) \nonumber\\
\{S^{+}(\rho)_{\a}^{\b}, S^{-}(\sg)_{\ga}^{\tau}\} &=& 0 \qquad \a,\b,\ldots =1,2. 
\label{pbs}
\eea
where $s$ is the intersection number (now set to $1$) and 
$\sg^{\ }_{1},\rho^{\ }_{1}$ (resp. $\sg^{\ }_{3},\rho^{\ }_{3}$) are the 
segments of paths {\it before} (resp. {\it after}) the intersection. 
The gauge invariance can be implemented by taking traces since, if $\da$ is an
open path
\be
R^{\pm}(\sg) = tr S^{\pm}(\sg) = tr S^{\pm}(\da^{-1}\sg \da) = 
R^{\pm}(\da^{-1}\sg \da)
\label{ttr}
\ee
where now $\da^{-1}\sg \da$ is closed. For $g = 1$ it is enough to have just six 
traces $~ R_i^{\pm}, i=1,2,3$, corresponding to the three paths $\ga_1, \ga_2 ,
\ga_3 = \ga_1\cdot\ga_2$. From \rref{pbs} they satisfy the non--linear cyclical 
Poisson bracket algebra \cite{NRZ}
\be
\{R_i^{\pm},R_j^{\pm}\}=\mp{\frac {\sqrt {-\Lambda}} 4}({\epsilon_{ij}}^k 
R_k^{\pm} - R_i^{\pm}R_j^{\pm}), \quad \epsilon_{123}=1
\label{pbt}
\ee 
and the cubic Casimir
\begin{equation}
1-(R_1^{\pm})^2-(R_2^{\pm})^2-(R_3^{\pm})^2 + 2 R_1^{\pm}R_2^{\pm}R_3^{\pm} = 0.
\label{b9}
\end{equation}
The traces (holonomies) of (\ref{pbt}) can be represented classically as
\be
R_1^\pm = \cosh r_1^\pm , \quad R_2^\pm  = \cosh r_2^\pm , \quad
R_3^\pm = \cosh(r_1^\pm+r_2^\pm) , \label{trep}
\ee
where
$r_{1,2}^{\pm}$ are real, global, time-independent (but undetermined) 
parameters which, from (\ref{pbt}) satisfy the Poisson brackets\footnote{
The parameters $r_{1,2}^{\pm}$ used here have been scaled by a factor of 
$\um$ with respect to previous articles}
\be\{r_1^\pm,r_2^\pm\}=\mp \frac{\sqrt{-\Lambda}}{4}, \qquad 
\{r_{1,2}^+,r_{1,2}^-\}= 0.
\label{pbr}
\ee
The above fully reduced system can be easily quantized either by replacing
the Poisson brackets \rref{pbr} by the commutators
\be
\left[{\hat r}_1^\pm,{\hat r}_2^\pm\right]=\mp \frac{i\hbar \sqrt{-\Lambda}}{4}, 
\qquad  \left[{\hat r}_{1,2}^+,{\hat r}_{1,2}^-\right]= 0.
\label{commr} 
\ee
or by directly quantizing the algebra \rref{pbt}. This gives for the
($+$) algebra\footnote{The ($-$) algebra is the same as \rref{commt} but uses
$q^{-1}$ rather than $q$} 
\be
q^{\um}{\hat R}_1^{+}{\hat R}_2^{+} 
- q^{-\um}{\hat R}_2^{+}{\hat R}_1^{+}
= (q^{\um} - q^{-\um}){\hat R}_3^{+}
\label{commt}
\ee 
where $q= \exp{2i\theta}$ and $\tan{\theta}=-\frac{\hbar \sqrt{-\Lambda}}{8}$. 
The algebra \rref{commt} is related \cite{NR2} to the Lie algebra of the quantum 
group $SU(2)_q$, and can be represented (up to rescalings of $O(\hbar)$) by e.g. 
${\hat R}_i = \um(A_i + A_i^{-1}), ~i=1,2,3$ where the $A_i$ satisfy 
\be
A_1 A_2 = q A_2 A_1, \qquad A_1 A_2 A_3 = q^{\um}
\label{a12}
\ee 
The first of \rref{a12} is called either a $q$--commutator, or a quantum
plane  relation, or it is said that $A_1, A_2$ form a Weyl pair. Relations
\rref{a12} can be satisfied by the assignments  
$A_1 = e^{{\hat r}_1}, A_2 =
e^{{\hat r}_2}, A_3 = e^{-({\hat r}_1 + {\hat r}_2)}$ 
with ${\hat r}_1, {\hat r}_2$ satisfying \rref{commr}.
\section{Classical and Quantum Equivalence}\label{sec5}
\subsection{Classical equivalence}\label{sec51}
The classical solution of Section \rref{metadm} can be related to the 
parameters $r_{1,2}^{\pm}$ of Section \rref{triadm} as follows \cite{cn}.

The ADM reduced actions \rref{ein3} and \rref{eintri2} are related
by
\bea
 I &=&~\int\!dt \int\!d^2x\,\pi^{ij} {\dot g}_{ij}=
 \int\!d^3x ~~{\Omega_i}^{ab}{{\dot{\Omega}}_j}^{CD}\ep_{ABCD}\nonumber\\
&&=~\int\um (\bar p dm + p d\bar m) -H dT -d(p^1m_1 +p^2m_2)\nonumber\\ 
&&=~\int \a(r_1^-dr_2^- - r_1^+ dr_2^+) 
\label{acts}
\eea
and show that with the time coordinate $t$ determined by 
$T=-\frac{2}{\a}\cot \frac{2t}{\a}$ the parameters $r_{1,2}^{\pm}$ are related 
to the complex modulus $m$ and momentum $p$ through a (time-dependent) 
canonical transformation. Explicitly, with 
$r_a(t) = r_a^-e^{\frac{it}{\a}} + r_a^+e^{-\frac{it}{\a}},~~ a=1,2.$ 
and the $r_a^{\pm}$ satisfying \rref{pbr}, then 
\be
m= {r_2}^{-1}(t){r_1(t)}, \qquad  \mathrm and \qquad 
p = -i \frac{\sqrt{T^2-4\La}}{4\La} {{\bar r}_2}{}^2 (t)
\label{mp} 
\ee
will satisfy the Poisson brackets \rref{pbtor}. 

The Hamiltonian \rref{hamtor} is now 
\be H=\frac{1}{\sqrt{T^2-4\Lambda}}(r_1^-r_2^+ - r_1^+r_2^-) 
\ee
and generates the development of the modulus and momentum \rref{mp} as functions
of the parameters $r_a{}^{\pm}$ and time $T$ through
\be
\frac{dp}{dT}=\{p,H\},\qquad \frac{dm}{dT}=\{m,H\} 
\label{hameq}\ee

\subsection{Large Diffeomorphisms }\label{sec52}
The reduction to the modulus $m$ and momenta $p$ means there are no more 
``small diffeomorphisms''- coordinate transformations (the constraints which 
generate them are all  identically zero). But there remain``large
diffeomorphisms'' due to the topology. These are transformations that are not
connected to the identity, cannot be built up from infinitesimal
transformations and are generated by ``Dehn twists'', i.e. by the operation of
cutting open a handle, twisting one end by $2\pi$, and regluing the cut edges. 
For $g>1$ the set of equivalence classes of such large diffeomorphisms (modulo
diffeomorphisms that can be deformed to the identity) is known as the mapping
class group.  For $g=1$ it is also called the modular group, and the Dehn
twists of the two independent circumferences $\ga_1$ and $\ga_2$ (which have
intersection number $+1$) act by
\begin{eqnarray}
&S&:\ga_1\rightarrow \ga_2^{-1},\hphantom{\ga_2}
    \qquad \ga_2\rightarrow\ga_1\nonumber\\
&T&:\ga_1\rightarrow\ga_1\cdot\ga_2 ,\qquad\ga_2\rightarrow\ga_2 ,
\label{modga}
\end{eqnarray}
These transformations induce the modular transformations  
\bea 
S:& m \to -m^{-1} \qquad &p \to {\bar m}^2 p\nonumber\\ 
T:& m \to m+1 \qquad &p \to p. 
\label{modm}
\eea
which preserve the Poincar\'e metric \rref{pmet}, the Hamiltonian \rref{hamtor} 
and the Poisson brackets \rref{pbtor}. The figure illustrates this group action 
on the modulus configuration space, with the invariant semicircle representing 
the geodesic motion of the modulus $m$.
\begin{figure}[ht]
\centerline{\epsfxsize=3.4cm\epsfbox{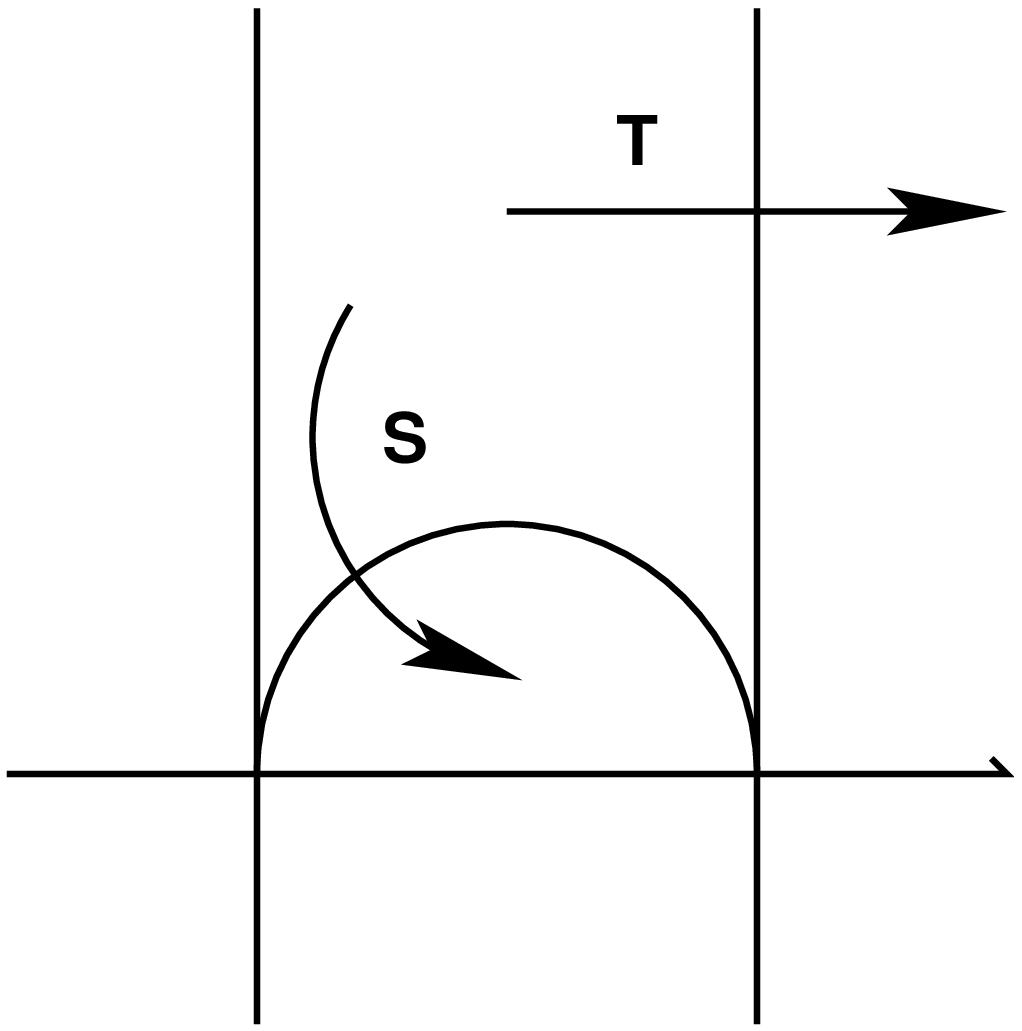}}   
\end{figure}

Classically, one could ask that observables be invariant under all spacetime
diffeomorphisms, including those in the modular group. Since equation
\rref{modm} shows that the modular group is well behaved on configuration
space, invariant functions of $m$ exist (see \cite{ord}). So the reduced ADM 
approach of Section \ref{metadm} looks like a standard ``Schr{\"o}dinger 
picture'' quantum theory, with time-dependent states $\psi(m,T)$ whose 
evolution is determined by the Hamiltonian operator \rref{hamop}. 

On the traces of holonomies the transformations \rref{modga} induce 
the following 
\begin{eqnarray}
S:& R_1^{\pm}\rightarrow  R_2^{\pm},
  \quad  R_2^{\pm}\rightarrow  R_1^{\pm},
  \quad  R_3^{\pm}\rightarrow
        2  R_1^{\pm}  R_2^{\pm} -  R_3^{\pm}\nonumber\\
T:& R_1^{\pm}\rightarrow  R_3^{\pm},\quad
   R_2^{\pm}\rightarrow  R_2^{\pm}, \quad
   R_3^{\pm}\rightarrow 2 R_3^{\pm} R_2^{\pm} - R_1^{\pm} .
\label{modt}
\end{eqnarray}
which preserve the algebra \rref{pbt}. The corresponding transformations on 
the holonomy parameters preserve their Poisson brackets \rref{pbr}  
\begin{eqnarray}
S&:~r_1^{\pm}\rightarrow r_2^{\pm}, \qquad \qquad
    &r_2^{\pm}\rightarrow - r_1^{\pm}\nonumber\\
T&:~r_1^{\pm}\rightarrow r_1^{\pm} + r_2^{\pm}, \qquad
    &r_2^{\pm}\rightarrow r_2^{\pm}.
\label{modr}
\end{eqnarray}
In this approach the modular group action \rref{modr} on the
parameters is {\it not} well behaved since it mixes $r_1$ and $r_2$, and
quantization normally requires a polarization. So the quantum theory of Section
\ref{triadm} resembles a ``Heisenberg picture'' quantum theory, with 
time-independent states $\psi(r)$, and, for some ordering, time--dependent 
operators \rref{mp}. 

\subsection{The Quantum Modular Group}\label{sec53}
Here I present work in collaboration with Carlip \cite{cn1}. It is useful to 
note that the modular transformations can also be implemented
quantum-mechanically, by conjugation with the unitary operators \cite{NR1,cn1}
\bea
&{\hat T}&= {\hat T}^+{\hat T}^-
= \exp\left\{ \frac{i}{2\hbar}({\hat p}+ {\hat p}^\dagger)\right\}\nonumber\\
&{\hat S}&= {\hat S}^+{\hat S}^-
= \exp\left\{ \frac{i\pi}{8\hbar} \left[
2({\hat p}^{\dag} + \hat p) + {\hat m}^{\dag}({\hat m}^{\dag}\hat p
+ \hat p {\hat m}^{\dag})
+ ({\hat m}{\hat p}^{\dag} + {\hat p}^{\dag} {\hat m}){\hat m}
\right]\right\}\nonumber
\eea
The $S$ transformation for $p$ differs from its classical version \rref{modm} 
\begin{equation}
S: \hat p\rightarrow  \frac{{\hat m}^{\dag}}{2}
({\hat m}^{\dag} \hat p + \hat p {\hat m}^{\dag}) ,
\label{s1}
\end{equation}
by terms of order $\hbar$. In terms of the holonomy parameters these are 
\begin{equation}
{\hat T}^{\pm} =
  \exp \left\{\pm \frac{i\alpha}{2\hbar}({\hat r}_2^{\pm})^2\right\} ,
\quad {\hat S}^{\pm} =
\exp \left\{\pm \frac{i\pi\alpha}{4\hbar}\left[({\hat r}_1^{\pm})^2
+ ({\hat r}_2^{\pm})^2 \right]\right\} ,
\label{t}
\end{equation}
Using the above construction the two representations, classically equivalent
as shown in Section \ref{sec51} can be related as follows. Start by
diagonalizing the commuting moduli operators $\hat m$ and ${\hat m}^\dagger$, 
considered
as functions of time and initial data $r_{1,2}^\pm$ through \rref{mp}). Now if
the $r_2(t)$ are ``coordinates'' $u(t)$ and $r_1(t)$ their ``momenta'' then
$\hat m$ and $\hat p$ act as
\begin{equation}
{\hat m} \sim u^{-1}\frac{\partial}{\partial u} \qquad
\hat p \sim {\bar u}^2
\label{pr}
\end{equation}
The normalized eigenstates of $\hat m$ with eigenvalues $m$ (and $\bar m$ for 
${\hat m}^\dagger$) are
\begin{equation}
K(m, \bar m,t|u,{\bar u})
  = \frac{\alpha m_2}{2\pi\hbar}{\bar u}\exp\left\{
  -\frac{\alpha}{4\hbar}m u^2
  + \frac{\alpha}{4\hbar}{\bar m}{\bar u}^2 \right\} .
\label{k}
\end{equation}
So candidates for ``Schr{\"o}dinger picture'' wave functions are the superpositions
\be
{\tilde\psi}(m,\bar m,t) =
  \int du_1du_2 K^*(m,\bar m,t|u,{\bar u})\psi(u,{\bar u}) 
\label{sch}
\ee
of the ``Heisenberg picture'' wave functions $\psi(u,\bar u)$. Inverting
\rref{sch} gives
\be 
\psi(u,{\bar u}) =
  \int_{{\mathcal F}} \frac{d^2 m}{m_2{}^2} K(m,\bar m,t|u,{\bar u})
  {\tilde\psi}(m,\bar m,t) .
\label{hei}
\ee 
where ${\mathcal F}$ is a fundamental region for the modular group. Now apply 
the $T$ transformation \rref{modm} to \rref{hei}
\bea  
{\hat T}\psi(u,{\bar u})
&=& {\hat T}\int_{{\mathcal F}}\frac{d^2m}{m_2{}^2}
K(m,\bar m,t|u,{\bar u}){\tilde\psi}(m,\bar m,t)\nonumber\\
&=& \int_{{\mathcal F}}\frac{d^2 m}{m_2{}^2}
K(m+1,\bar m +1,t|u,{\bar u}){\tilde\psi}(m+1,\bar m +1,t)
\nonumber\\
&=&\int_{T^{-1}{\mathcal F}}
 \frac{d^2m }{m_2{}^2} K(m,\bar m,t|u,{\bar u})
 {\tilde\psi}(m,\bar m,t) ,
\label{thei}
\end{eqnarray}
where $T^{-1}{\mathcal F}$ is the new fundamental region obtained from 
$\mathcal F$ by a $T^{-1}$ transformation, and in \rref{thei}
${\tilde\psi}(m,\bar m,t)$ and the integration measure are modular invariant. A
similar argument holds for the $S$ transformation, and shows that there are
no  invariant ``Heisenberg picture'' wave functions  $\psi(u,{\bar u})$, since
the integration regions in \rref{hei} and \rref{thei} are disjoint except on a set
of measure zero. Further, $\psi(u,{\bar u})$ and ${\hat T}\psi(u,{\bar u})$ 
are orthogonal since
\begin{equation}
\left\langle \psi|{\hat T}\psi\right\rangle =
\int_{T^{-1}{\mathcal F}}\frac{d^2m}{m_2{}^2}
\int_{{\mathcal F}}\frac{d^2m'}{m'_2{}^2}
m'_2{}^2 \delta^2(m-m')
{\tilde\psi}(m,{\bar m},t){\tilde\psi}^*(m',{\bar m}',t)=0 ,
\label{matel}
\end{equation}
and similarly for $S$.

Equation (\ref{matel}) shows that the modular group splits the Hilbert space of
square-integrable functions of $(u_1,u_2)$ into an infinite set of orthonormal
fundamental subspaces consisting of wave functions of the form \rref{hei} for
a fixed fundamental region $\mathcal F$. It is shown in \cite{cn1} that they
are physically equivalent, because matrix elements of invariant operators can
be computed in any of these subspaces, and each one is equivalent
to the ADM Hilbert space.

\end{document}